\documentclass[aps,prl,twocolumn,superscriptaddress]{revtex4-1}
\usepackage{graphicx,color}
\usepackage{amsmath,amssymb}

\begin{document}

\title{Inverse Problem Instabilities in Large-Scale Plasma Modelling}

\author{M.F. Kasim}
\author{T.P. Galligan}
\author{J. Topp-Mugglestone}
\author{G. Gregori}
\author{S.M. Vinko}
\email{sam.vinko@physics.ox.ac.uk}
\affiliation{Department of Physics, Clarendon Laboratory,University of Oxford, Parks Road, Oxford OX1 3PU, UK}

\date{\today}

\begin{abstract}
Our understanding of physical systems generally depends on our ability to match complex computational modelling with measured experimental outcomes. However, simulations with large parameter spaces suffer from inverse problem instabilities, where similar simulated outputs can map back to very different sets of input parameters. While of fundamental importance, such instabilities are seldom resolved due to the intractably large number of simulations required to comprehensively explore parameter space. Here we show how Bayesian machine learning can be used to address inverse problem instabilities, and apply it to two popular experimental diagnostics in plasma physics. We find that the extraction of information from measurements simply on the basis of agreement with simulations is unreliable, and leads to a significant underestimation of uncertainties. We describe how to statistically quantify the effect of unstable inverse models, and describe an approach to experimental design that mitigates its impact.
\end{abstract}

\pacs{Valid PACS appear here}
\maketitle

Our understanding of physical systems is closely dependent on our ability to predictively model their behaviour in controlled experimental settings. By conducting simulations which produce outcomes resembling those observed experimentally we can investigate which processes are important, and quantify their effect on observed outcomes. This approach is straightforward when experiments can be designed to probe a specific process in isolation, and for which the modelling contains as few variable parameters as possible. Unfortunately, few problems lend themselves to such scrutiny, and it is common for experimental outcomes to intertwine a host of competing processes, and for simulations to operate in large parameter spaces.
 An example of such research is the study of matter at high energy densities, i.e., systems at temperatures exceeding $\sim$10,000~K at the typical density of a solid, or at pressures exceeding 1~Mbar. Matter in these conditions is of great interest to astrophysical and inertial confinement fusion (ICF) investigations, but it tends to be highly transient, inhomogeneous, challenging to controllably create and difficult to probe. Research here relies strongly on complex computational modelling that needs to account for a wide range of processes and interactions, which, critically, depend on a large number of variable parameters. Some difficulties with this approach for inertial fusion energy research has been discussed within the context of insufficiently accurate models and diagnostics~\cite{Hinkel2013,Hurricane:2014aa}, but little attention has been paid to the intrinsic limitation of integrated experiments in their own right, and to the systematic uncertainties introduced by correlated physical parameters in both experiment and simulation.

Here we use Bayesian inference to explore the behaviour of complex, multi-parameter simulations, employing Markov Chain Monte Carlo (MCMC) algorithms to efficiently sample large dimensional spaces~\cite{Andrieu2003}. While our approach is general, we will focus specifically on two widely used experimental plasma diagnostics: x-ray spectroscopy and inelastic x-ray scattering. As we will show, inverse problem instabilities are significant in both cases, and must be comprehensively treated if meaningful information is to be extracted robustly from experimental measurements.

\section{Searching the Space of Solutions}

Simulations in physics are forward models: an initial set of parameters is chosen together with a set of models and assumptions, and is fed into an algorithm to calculate a set of outputs. The outputs typically correspond to the experimental observables, while the input parameters are related to the experimental conditions, or govern the models deployed. The models represent our best current knowledge of the physical system. The comparison between experiment and simulation is then the process by which we vary the inputs of the simulation to find outputs that match the experimental observables. Within this context, a simulation $f$ can be viewed as a map from a set of $n$ input parameters to a set of $m$ output observables $f: \mathbb{R}^n \rightarrow \mathbb{R}^m$.
The comparison with experiment is then simply the search for its inverse, $f^{-1}$, the function that maps the known experimental outputs $\mathbf{y} \in \mathbb{R}^m$ onto some vector of input parameters $\mathbf{x} \in \mathbb{R}^n$. If the number of input parameters is large, this inversion problem cannot be efficiently solved via a brute-force grid-search approach. This is due to the dimensionality curse: the number of simulations required to fill parameter space grows exponentially with the number of dimensions, quickly rendering the problem computationally intractable. The common workaround is to attempt to artificially reduce the range and number of parameters and restrict the search to a smaller space. However, this can introduce significant bias to the analysis. The problem of parameter bias is most acute if $f$ is not injective, i.e., when a point in the space of outcomes can map on to several possible values in the space of parameters. However, the problem remains relevant even for injective maps in the presence of experimental uncertainties, such as noise.

If similar sets of outcomes can be generated from very different sets of inputs, the inverse function $f^{-1}$ is said to be unstable: the inversion can yield multiple solutions, indistinguishable within the experimental uncertainty. Nevertheless, the problem can be treated statistically, by searching for the likelihood $P$ of finding a specific set of parameters $\mathbf{x}$ given an observed outcome $\mathbf{y}$, $P(\mathbf{x} | \mathbf{y})$. Finding this probability can therefore be phrased as a Bayesian inference problem~\cite{Berger:1985}:
\begin{equation}
P (\mathbf{x} | \mathbf{y}) = \frac{P (\mathbf{y} | \mathbf{x}) P(\mathbf{x})}{P (\mathbf{y})},
\end{equation}  
where $P(\mathbf{y} | \mathbf{x})$ is the likelihood of finding the observables $\mathbf{y}$ given a set of input parameters $\mathbf{x}$ (a forward model calculation), $P(\mathbf{x})$ is the prior distribution of possible parameters, and $P(\mathbf{y})$ is the marginal likelihood of the observed data over all possible parameters. The prior distribution here is important as it provides a way to bias the parameters due to physical constraints, or to use results from complementary diagnostics, the results of which would otherwise not feed into the evaluation of the observed output. The marginal likelihood $P (\mathbf{y})$ is intractable to compute in high dimensional space, but is simply a scaling constant given its independence on $\mathbf{x}$. We chose the forward model likelihood $P(\mathbf{y} | \mathbf{x})$ to be uniform if the simulated output lies within some uncertainty band around the data we aim to reproduce, and zero otherwise.

To sample the posterior distribution of the parameters, $P(\mathbf{x} | \mathbf{y})$, we employ Markov Chain Monte Carlo (MCMC) algorithms~\cite{Metropolis:1953,Hastings:1970,Goodman:2010} to obtain samples from an unknown probability distribution in high dimensional spaces~\cite{Andrieu2003}.
This approach provides an efficient way to sample the possible simulations that yield outcomes which match, to within a given uncertainty, the experimentally observed result. Importantly, while all matching simulations have equal prior probability, their posterior probability distribution in parameter space will not generally be uniform. We can thus address the inverse problem instability by investigating the posterior probability distribution of parameters and give best-estimates that are statistically meaningful even for highly pathological, non-invertible forward models.

\subsection{Inelastic x-ray scattering}

Inelastic x-ray Thomson scattering (XRTS) is a widely used tool in high energy density physics to determine the conditions of warm-dense matter via the direct probing of the electron response function~\cite{Glenzer:2009,PhysRevE.67.026412,Garcia-Saiz:2008aa}. This includes measuring plasma temperatures and densities~\cite{Kritcher69,PhysRevLett.108.145001}, ion correlations~\cite{Barbrel:2009,PhysRevLett.110.065001}, transport properties~\cite{Faustlin:2010,PhysRevLett.115.115001}, ionization and continuum lowering~\cite{Fletcher2013,PhysRevLett.112.145004}, and informing equation of state studies~\cite{Lee:2009,PhysRevLett.109.265003,Fletcher:2015aa,PhysRevE.94.011202}.
In XRTS, the backscattering spectrum provides access to the non-collective Compton scattering regime which can probe the temperature, density, and ionization state of the plasma, while the forward scattering spectrum is sensitive to collective plasmon oscillations.

\begin{figure}
\includegraphics[width=\linewidth]{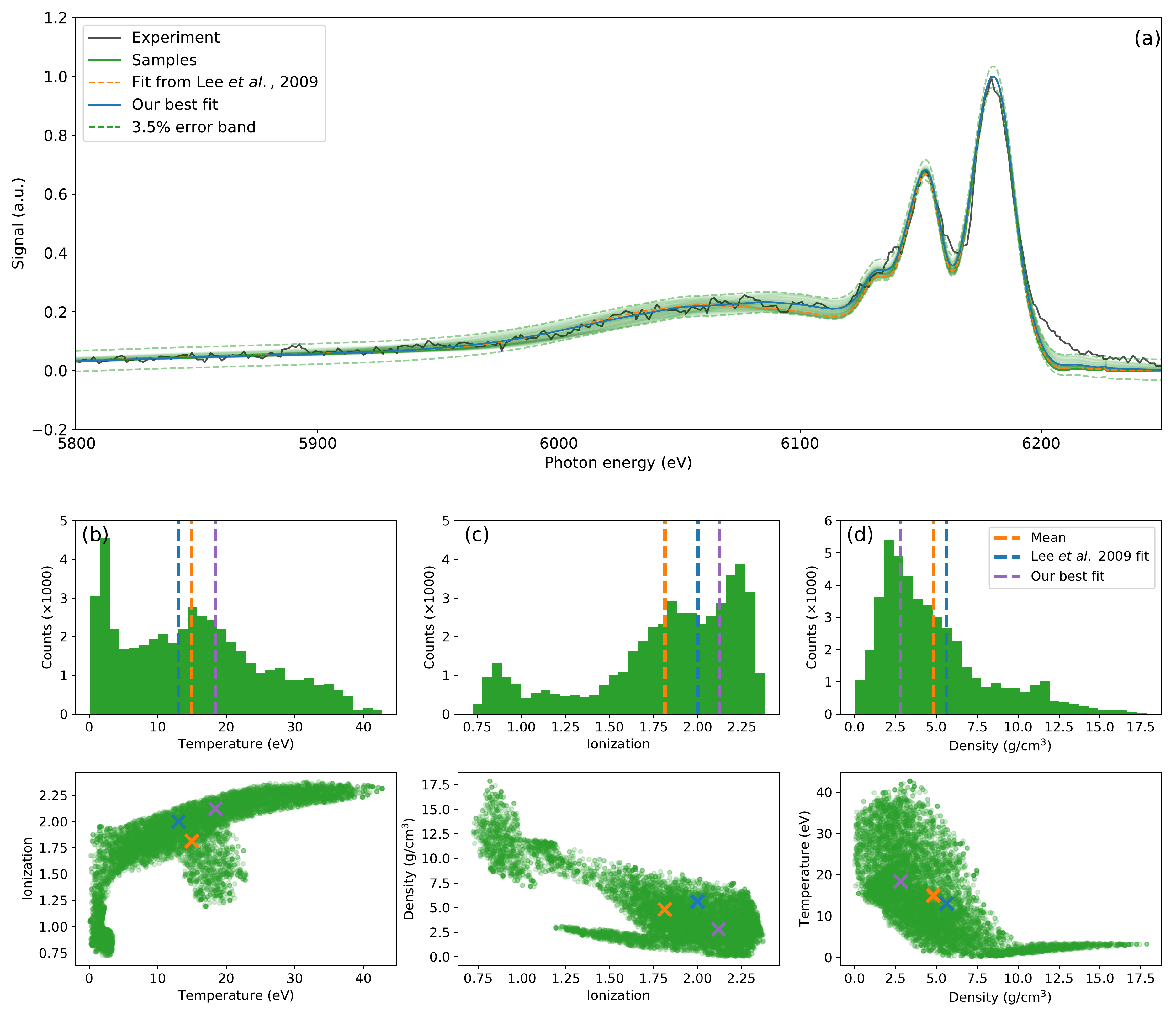}
\caption{\label{FIG-XRTS} {\bf X-ray Thomson scattering:} (a) Experimental and fitted spectrum (from ref.~\cite{Lee:2009}) compared with our simulated sample spectra with total deviations below 3.5\% (green).
(b-d) Posterior distribution histograms for the deduced temperature, density, and ionization using a hard-boundary forward likelihood, and assuming the prior distribution $P(\mathbf{x})$ is uniform for all parameters. We plot out best fit and average parameter value, alongside the original conditions from ref.~\cite{Lee:2009}.}
\end{figure}

To explore the inverse problem instability in XRTS we use the model of Gregori {\it et al.}~\cite{Gregori:2007a,PhysRevE.74.026402} within the Chihara approximation~\cite{Chihara:1987}. This is a forward model that takes as input a set of plasma parameters (electron and ion temperature, ionization, density, spectrum of incident x-rays), and produces an energy-resolved x-ray scattering spectrum. The scattering spectrum is then convolved with the point spread function of the spectrometer to predict the experimental observation.

Our analysis is generally applicable to any XRTS experiment, but to focus our discussion and derive quantitative results we will analyse the high-quality experimental data of Lee \textit{et al}.~\cite{Lee:2009}. In this experiment, XRTS was used to deduce the temperature and density of shock-compressed beryllium. We show the scattering spectrum from ref.~\cite{Lee:2009} in Fig.~\ref{FIG-XRTS}a, alongside a 3.5\% error band within which all our simulated scattering spectra are required to fall. The full set of collected simulated spectra obtained by our MCMC algorithm is given in Fig.~\ref{FIG-XRTS}b for the electron temperature, mass density, and ionization. We find a large parameter spread ranging orders of magnitude in temperature and density, all while yielding total scattering spectra that are indistinguishable within the noise. This large spread illustrates the difficulties that arise in data interpretation due to the instability of the inverse problem, as any point could potentially represent the experiment.

For each point in Fig.~\ref{FIG-XRTS} we know, by construction, that the probability $P(\mathbf{y} | \mathbf{x})$ of finding the experimental spectrum given the parameters (density, temperature and ionization) is large. However, this alone does {\it not} imply that the observed spectrum favours that specific set of input parameters, i.e., that $P(\mathbf{x} | \mathbf{y})$ is also large. In fact, the posterior distribution probability $P(\mathbf{x} | \mathbf{y})$ is generally unknown unless calculated explicitly. We show our calculations for the posterior parameter distributions for this case in Fig.~\ref{FIG-XRTS} (b-d).

Not considering posterior distributions in interpreting the data can lead to contradictions and erroneous conclusions. For example, a prior assumption that the temperature should be around 40~eV would imply a corresponding density of around 3.4~gcm$^{-3}$, and an ionization of $Z\sim2.3$. However, assuming the temperature to be around 0.15~eV would yield a density estimate in excess of 7~gcm$^{-3}$ and an ionization of 1. Note that both of these contradictory estimates rely on comparable agreement with experimental data. This exemplifies the insidious problem where prior assumptions -- the bias of the modeller -- can completely determine outcomes, irrespective of the underlying experimental data. It is therefore important to evaluate the posterior distribution probabilities rather than simply relying on agreement between simulation and experiment.
The posterior distribution allows us to quantify which plasma conditions are most likely, and to estimate uncertainties. From the histograms in Fig.~\ref{FIG-XRTS} we find mean values to be $T=15^{+21}_{-14}$~eV, $Z=1.8^{+0.5}_{-1.0}$ and $\rho = 5^{+8}_{-4}$~gcm$^{-3}$. The uncertainties quoted are 95\% confidence intervals, and are significantly larger than those originally reported.

\subsection{X-ray spectroscopy}

\begin{figure}
\includegraphics[width=\linewidth]{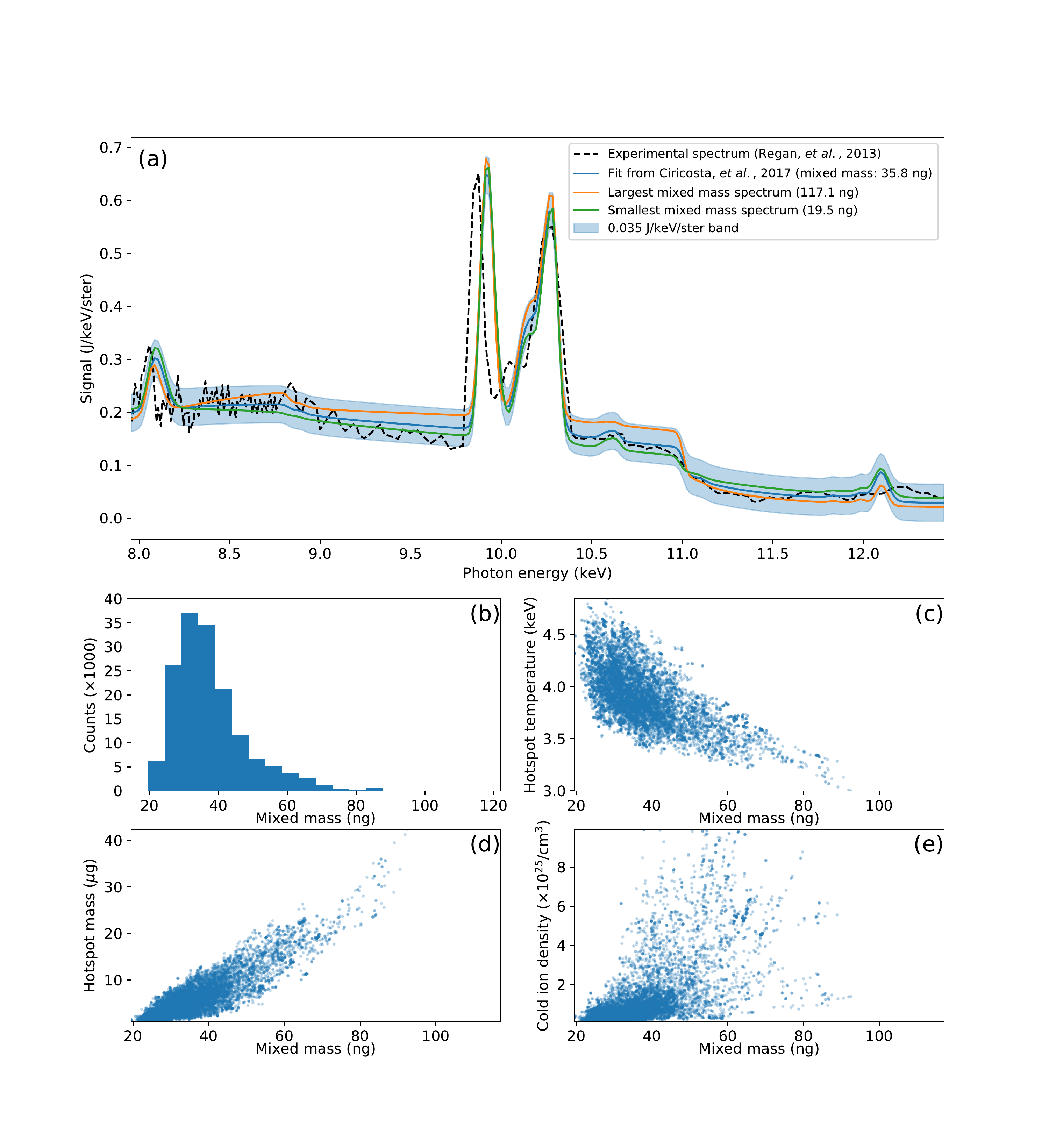}
\caption{\label{FIG-spec} {\bf X-ray emission spectroscopy:} (a) Emission spectrum from a NIF implosion experiment (from ref.~\cite{Regan:2013}), alongside the simulated spectrum from ref.~\cite{Ciricosta:2017}, and two simulations from this work that yield the largest and smallest mixed masses. The band represents the experimental uncertainty. Simulation outcomes from the MCMC algorithm in 10-D parameter space showing in (b) the posterior distribution of the mixed mass, and in (c-e) how it depends on hotspot mass, temperature and cold density.}
\end{figure}

X-ray spectroscopy is a popular diagnostic technique across the physical sciences, widely used in astrophysical investigations~\cite{Labrosse2010,Cottam:2002aa,Bailey:2014aa}, for elemental analysis in laser-induced breakdown spectroscopy~\cite{Hahn:2010}, in applied condensed matter physics \cite{Liu:2013aa,PhysRevLett.100.167402}, in fundamental plasma physics~\cite{Vinko2012,quincy:2018}, laser-plasma interactions~\cite{PhysRevA.46.R1747,PhysRevLett.110.125001,Hoarty2013}, and in inertial confinement fusion (ICF) research~\cite{Regan:2002,PhysRevLett.70.1263,Regan:2013,PhysRevLett.111.085004}. 

\begin{figure*}
\includegraphics[width=\linewidth]{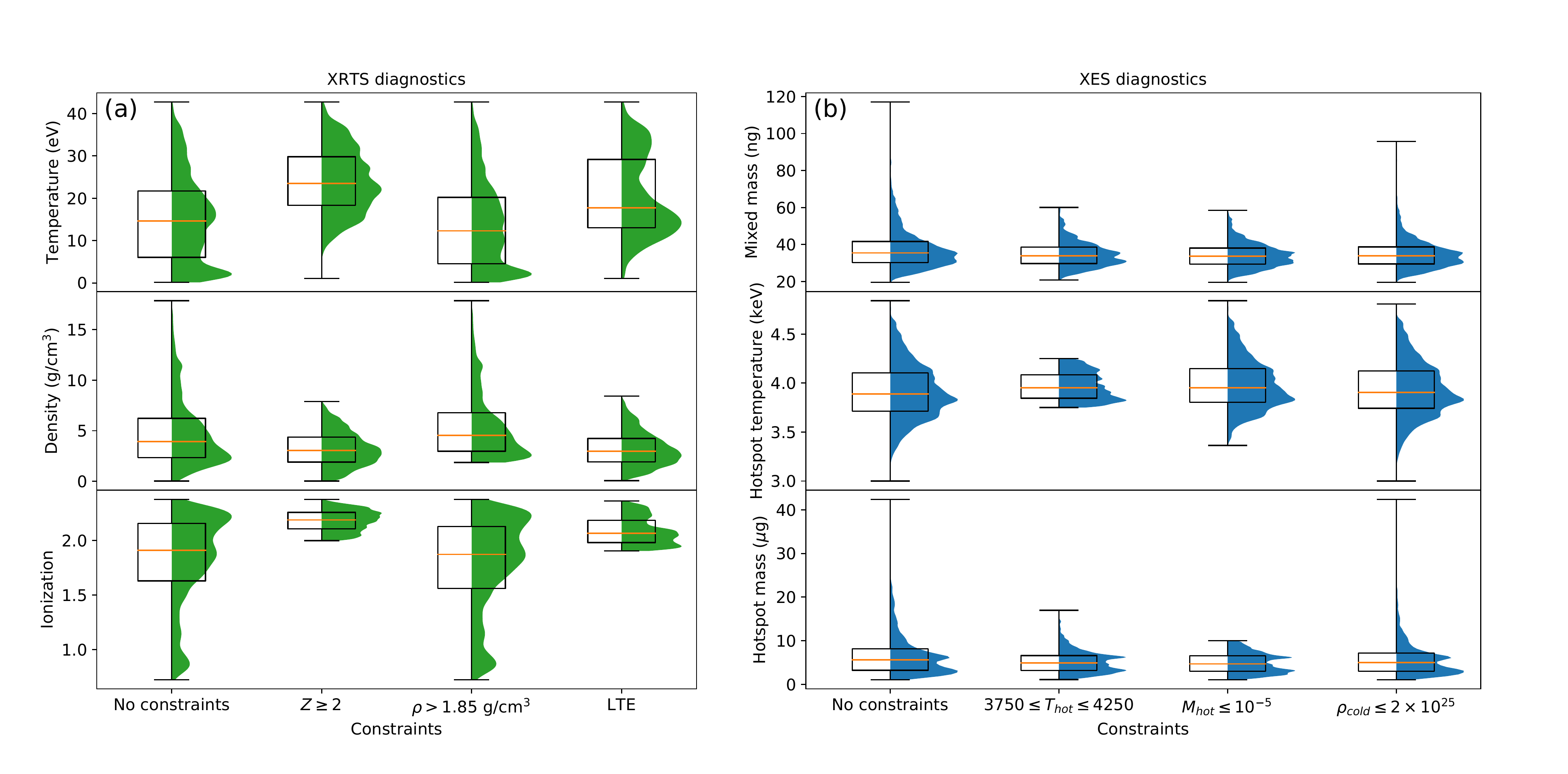}
\caption{\label{FIG-errorbars}  {\bf Effect of constraints:} Posterior probability distributions with various constraints for (a) x-ray Thomson scattering and (b) x-ray emission spectroscopy. The boxes represent 50\% confidence intervals and show the median value. The vertical bars indicate the full range of possible parameters.}
\end{figure*}

To explore the role of inverse problem instabilities in the interpretation of spectroscopic data we will examine emission spectroscopy used in ICF experiments on the National Ignition Facility (NIF)~\cite{Lindl:2004,Regan:2013}. Specifically, it was observed in these experiments that the mixing of cold material into the hot spot of an imploding capsule via hydrodynamical instabilities can quench the ignition process by enhancing radiative and conductive losses~\cite{Hammel:2010}. It is thus of considerable interest to understand the mechanisms and amount of mix taking place in experiments. To this end, a novel spectroscopic model was proposed by Ciricosta {\it et al.}~\cite{Ciricosta:2017} that makes use of the spectroscopic emission signature to infer the amount of mixed mass. Here, the emission was simulated by the atomic-kinetics radiation-transfer code Cretin~\cite{Scott:2001}, extensively benchmarked to large-scale hydrodynamic simulations. The simulation is a forward model that takes as input ten parameters describing the various parameters of the imploding capsule, including a jet of cold material driven towards the core, and produces a synthetic x-ray emission spectrum that can be compared with experiment. A comprehensive description of the ten parameters and their significance is beyond the scope of this paper, but can be found in ref.~\cite{Ciricosta:2017}. For the purpose of the present discussion, we will focus only on trying to infer the mass of cold material that mixes into the hotspot (mixed mass) by searching for matching simulated and experimental emission spectra.

As previously, we use MCMC to calculate posterior distributions of the modelling parameters, by using Ciricosta's forward model to simulate the experimental data of Regan \textit{et al.}~\cite{Regan:2013}. We again assume a uniform distribution of priors.
The emission spectra are shown in Fig.~\ref{FIG-spec}a, where we compare the experimental data with the fitting result and deduced mixed mass from ref.~\cite{Ciricosta:2017}, alongside two of our calculations that yield the largest and smallest mixed mass estimates. All simulations are contained within the experimental error bars.
Regan \textit{et al.} initially estimated a mixed mass of $34^{+50}_{-13}$~ng by fitting only the Ge He$_{\alpha}$ complex, while Ciricosta {\it et al.} estimated a mixed mass of 35~ng by fitting to the whole spectrum shown in Fig.~\ref{FIG-spec}, estimating their uncertainty to be within $\sim$80\%.
In contrast, we find spectra that match the experimental data with associated values of the mixed mass in the entire range between 20-117~ng.
Nevertheless, as shown in Fig.\ref{FIG-spec} (b), our posterior probability distribution for the mixed mass exhibits a distinct peak around 35~ng, a value largely in line with previous estimates. Specifically, we find $38^{+27}_{-14}$~ng, with a 95\% confidence interval. So while a wide range of mix values are consistent with the data, certain values are more probable than others.
This observation seems encouraging for the exploitation of x-ray emission spectroscopy to measure mixed mass in ICF experiments, despite the intrinsically large uncertainties of the approach. Estimates made during the National Ignition Campaign place a mixed-mass limit of around 75~ng for a hotspot of $\sim$2$\times 10^4$~ng before significant implosion efficiency deterioration~\cite{Haan:2011}. Diagnostics are therefore needed that are accurate to well within that range, and this evaluation of x-ray emission spectroscopy suggests the inversion could be made robust to the required level.

\section{The effect of constraints}

So far we have assumed there were no constraints on the parameters used for the simulations. However, complementary experimental diagnostics and other limitations can constrain the possible space of parameters to be explored. This is equivalent to adding a prior bias to the parameter set. It is thus interesting to see how constraints can influence posterior distribution probabilities, and by them our conclusions on the measured outcomes.

We illustrate the effect of three different types of constraints on the outcome of a XRTS measurement in Fig.~\ref{FIG-errorbars} (a). The panels show the distributions of the deduced temperature, density and ionization in the Be plasma, with the respective median value. The constraint that the ionization of Be is always larger than 2 strongly affects the range of possible densities, and noticeably raises the estimated temperature. We note that in the original work the ionization was set to $Z=2$ on the basis of radiation-hydrodynamic modelling~\cite{Lee:2009}. Such an assumption is seen to have a profound effect on the interpretation of the experimental results and predetermines the range of possible outcomes. In contrast, the constraint that the Be is compressed above solid density is seen not to influence the results, as most simulations that match the data already require higher densities. Given the timescales of the experiment one could also suppose the Be system to be near local thermodynamic equilibrium (LTE). Under this assumption the range of densities and ionisations again decreases, but leads only to a modest change in temperature.

The effects of parameter constraints in inverting x-ray spectroscopy models are shown in Fig.~\ref{FIG-errorbars} (b). Here we assume that complementary diagnostics in an ICF implosion experiment are able to limit the possible variation of the temperature or mass of the hot spot, or of the cold density.
Such additional data, for example obtained from a neutron intensity measurement, can then be integrated in the parameter search by restricting the parameter space in the implosion model. As we can see, the expectation value of the mixed mass seems robust to the addition of constraints to the MCMC calculation and the uncertainties are lowered by over a factor two, provided the hotspot temperature can be determined to within $\sim$10\%. Our approach thus allows us not only to integrate various diagnostic outputs into the evaluation of specific observables, but also to estimate what additional parameters need to be measured, and with what accuracy, for the spectroscopic measurement to be robust.

\begin{figure}
\includegraphics[width=\linewidth]{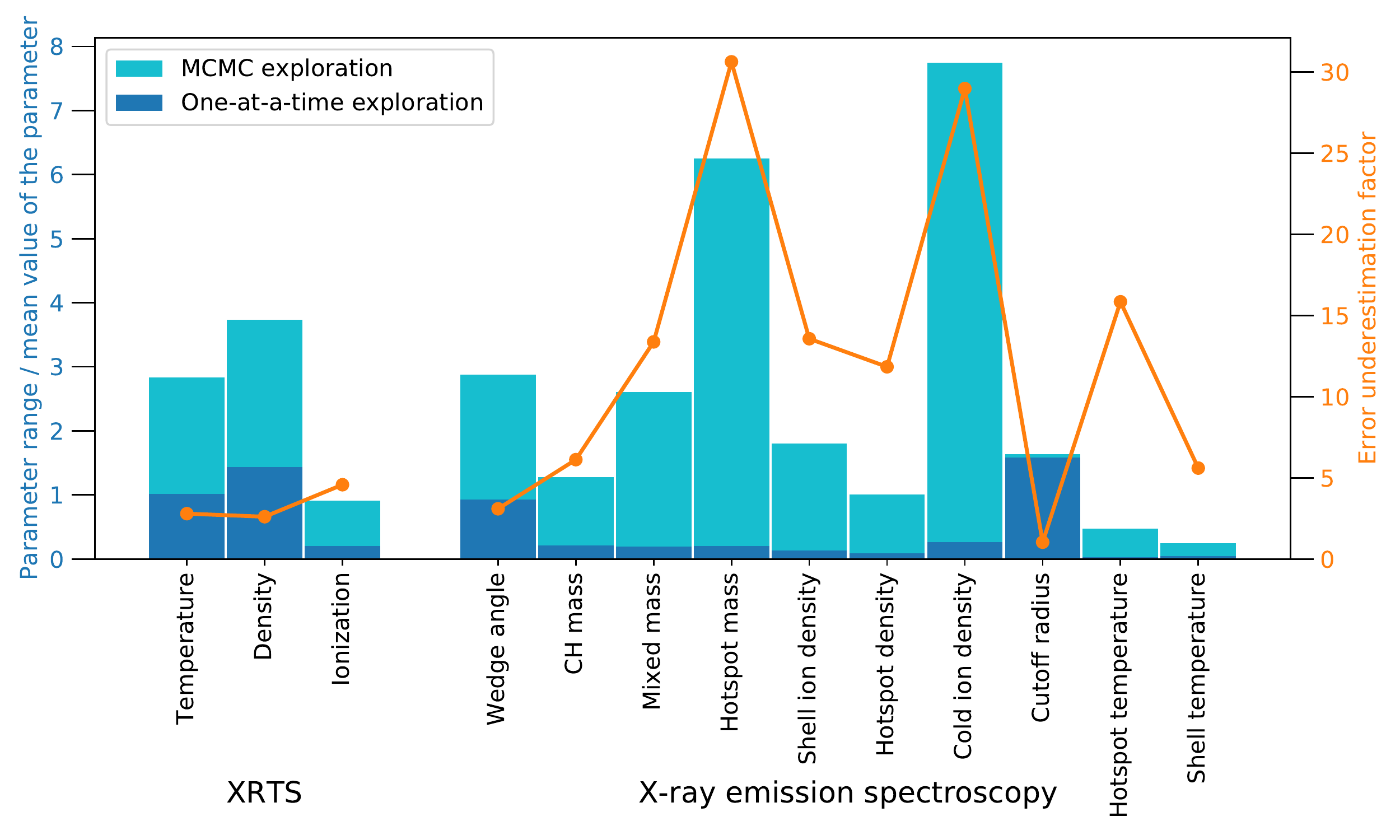}
\caption{\label{FIG-exploration} {\bf Uncertainty estimates:} Parameter uncertainties estimated via the full MCMC search of parameter space, compared with manually adjusting one parameter at the time, for the scattering (XRTS) and emission spectroscopy measurements. The single-parameter variation approach grossly underestimates the uncertainty, by up to a factor 30.}
\end{figure}

We note that the uncertainties quoted in the literature are generally much smaller than those deduced via our MCMC approach, often by over an order of magnitude. To understand why this happens we plot in Fig.~\ref{FIG-exploration} the deduced uncertainties for the two cases examined here, obtained in two different ways. The first is taken directly from the posterior distribution shown in Figs.~\ref{FIG-XRTS} and \ref{FIG-spec}. The second is based on the smallest variation we can make to a single parameter to produce simulations that no longer agree with the data to within the specified uncertainty -- the simplest way of estimating uncertainties in fits. We note that the variation of only a single parameter significantly underestimates the uncertainty, as the other parameters are not allowed to compensate for the changes made.

Performing experiments where some parameters can be explicitly constrained by the setup or by the interaction itself is clearly very appealing in the context of trying to minimise difficulties related to inverting physical models. For example, recent scattering measurements of the electrical conductivity in warm dense aluminium took place on time scales so short that the ion density remained frozen, thus removing all uncertainty from that parameter in the inversion~\cite{PhysRevLett.115.115001}. A similar frozen-ion constraint was also exploited in spectroscopic investigations of continuum lowering~\cite{Ciricosta2012,Ciricosta:2016aa}. Other experiments have instead added additional diagnostics such as x-ray diffraction to help place stringent experimental constraints on regions of the parameter space~\cite{Garcia-Saiz:2008aa,Fletcher:2015aa}.
However, such mitigation-by-design is not always possible, and some investigations suffer more strongly from inverse problem instabilities. Because the data with the highest levels of uncertainty will be most affected, experimental results with large error bands should be treated with caution, especially if used as a benchmark for physical properties such as plasma correlations~\cite{PhysRevLett.110.065001}, or in determining the equation of state of extreme states of matter~\cite{PhysRevE.94.011202,PhysRevLett.109.265003}.

For the cases discussed here we deliberately used small uncertainties to constrain the parameters, and have neglected several additional sources of uncertainty
For example, we have ignored the presence of gradients in the samples under investigation and the possibility that the measured outcome includes contributions from systems under very different conditions~\cite{Kozlowski:2016aa}. We have also ignored the effect of both model inaccuracy (e.g., in collision ionization models in atomic kinetics modelling~\cite{PhysRevLett.115.115001}) and model inadequacy (e.g., breakdown of the Chihara decomposition in XRTS~\cite{Crowley:2014aa}). Clearly, such effects can vary significantly across experiments, but their combined effect will further restrict the information that can reliably be extracted from the measurement. Nevertheless, provided that limits can be placed on these uncertainties, their effect on the posterior probability distributions can be investigated comprehensively within the present framework.

\section{Conclusions}

We have presented a study of inverse problem instabilities in complex simulations with large parameter spaces, applied to the interpretation of plasma physics experiments. We find that spectroscopic and scattering diagnostics can suffer from inverse instabilities to a significant extent, and that despite much effort in the field, can only lead to robust interpretations if the posterior probability distributions are considered, and in the presence of parameter constraints. The uncertainties arising from inverse instabilities are also significantly larger that those commonly quoted in the literature. Importantly, fielding appropriate complementary experimental diagnostics can be highly beneficial in restricting the possible space of solutions. In this regard, our Bayesian approach can provide valuable guidance in diagnostics selection, and on experimental design more generally.
While we have focused here on specific plasma physics experiments, we expect similar difficulties to affect all fields of science where large-scale, multi-parameter inverse models are used. The inverse problem instability is a common aspect of data modelling, and although it is seldom treated systematically, it can significantly alter the conclusions drawn from experiments. In this context, novel machine learning algorithms and approaches from data science can prove invaluable in guiding a range of research efforts, and in robustly quantifying experimental uncertainties while minimising unwanted bias.

\section{Acknowledgements}
We thank Dr S.~Regan at the Laboratory for Laser Energetics (University of Rochester, USA) for providing support for J. T-M. XRTS research.
S.M.V. is grateful for support from the Royal Society.
M.F.K. and S.M.V. acknowledge support from the UK EPSRC grant EP/P015794/1.
T.G. acknowledges support from the Oxford University Centre for High Energy Density Science (OxCHEDS).
G.G. acknowledges support from AWE plc., and the UK EPSRC (EP/M022331/1 and EP/N014472/1).


\begin{thebibliography}{10}

\bibitem{Hinkel2013}
D.~E. Hinkel et~al.
\newblock Progress toward ignition at the national ignition facility.
\newblock {\em Plasma Physics and Controlled Fusion}, 55(12):124015, 2013.

\bibitem{Hurricane:2014aa}
O.~A. Hurricane et~al.
\newblock Fuel gain exceeding unity in an inertially confined fusion implosion.
\newblock {\em Nature}, 506:343--348, 2014.

\bibitem{Andrieu2003}
C.~Andrieu, N.~de~Freitas, A.~Doucet, and M.I. Jordan.
\newblock An introduction to mcmc for machine learning.
\newblock {\em Machine Learning}, 50(1):5--43, 2003.

\bibitem{Berger:1985}
J.~O. Berger.
\newblock {\em Statistical Decision Theory and Bayesian Analysis}.
\newblock Springer, 2nd edition, 1985.

\bibitem{Metropolis:1953}
N.~Metropolis, A.~W. Rosenbluth, M.~N. Rosenbluth, A.~H. Teller, and E.~Teller.
\newblock Equation of state calculations by fast computing machines.
\newblock {\em J. Chem. Phys.}, 21(6):1087--1092, 1953.

\bibitem{Hastings:1970}
W.~K. Hastings.
\newblock Monte carlo sampling methods using markov chains and their
  applications.
\newblock {\em Biometrika}, 57(1):97--109, 1970.

\bibitem{Goodman:2010}
J.~Goodman and J.~Weare.
\newblock Ensemble samplers with affine invariance.
\newblock {\em Comm. App. Math and Comp. Sci.}, 5(1):65--80, 2010.

\bibitem{Glenzer:2009}
S.~Glenzer and R.~Redmer.
\newblock X-ray thomson scattering in high energy density plasmas.
\newblock {\em Reviews of Modern Physics}, 81:1625--1663, 2009.

\bibitem{PhysRevE.67.026412}
G.~Gregori, S.~H. Glenzer, W.~Rozmus, R.~W. Lee, and O.~L. Landen.
\newblock Theoretical model of x-ray scattering as a dense matter probe.
\newblock {\em Phys. Rev. E}, 67:026412, 2003.

\bibitem{Garcia-Saiz:2008aa}
E.~Garc{\'\i}a~Saiz et~al.
\newblock Probing warm dense lithium by inelastic x-ray scattering.
\newblock {\em Nature Physics}, 4:940--944, 10 2008.

\bibitem{Kritcher69}
Andrea~L. Kritcher et~al.
\newblock Ultrafast x-ray thomson scattering of shock-compressed matter.
\newblock {\em Science}, 322(5898):69--71, 2008.

\bibitem{PhysRevLett.108.145001}
A.~J. Visco et~al.
\newblock Measurement of radiative shock properties by x-ray thomson
  scattering.
\newblock {\em Phys. Rev. Lett.}, 108:145001, 2012.

\bibitem{Barbrel:2009}
B.~Barbrel et~al.
\newblock {Measurement of Short-Range Correlations in Shock-Compressed Plastic
  by Short-Pulse X-Ray Scattering}.
\newblock {\em Phys. Rev. Lett.}, 102(16):2--5, 2009.

\bibitem{PhysRevLett.110.065001}
T.~Ma et~al.
\newblock X-ray scattering measurements of strong ion-ion correlations in
  shock-compressed aluminum.
\newblock {\em Phys. Rev. Lett.}, 110:065001, 2013.

\bibitem{Faustlin:2010}
R.~R. F\"austlin et~al.
\newblock Observation of ultrafast nonequilibrium collective dynamics in warm
  dense hydrogen.
\newblock {\em Phys. Rev. Lett.}, 104:125002, Mar 2010.

\bibitem{PhysRevLett.115.115001}
P.~Sperling et~al.
\newblock Free-electron x-ray laser measurements of collisional-damped plasmons
  in isochorically heated warm dense matter.
\newblock {\em Phys. Rev. Lett.}, 115:115001, Sep 2015.

\bibitem{Fletcher2013}
L.~B. Fletcher et~al.
\newblock X-ray thomson scattering measurements of temperature and density from
  multi-shocked ch capsules.
\newblock {\em Phys. Plasmas}, 20(5):056316, 2013.

\bibitem{PhysRevLett.112.145004}
L.~B. Fletcher et~al.
\newblock Observations of continuum depression in warm dense matter with x-ray
  thomson scattering.
\newblock {\em Phys. Rev. Lett.}, 112:145004, Apr 2014.

\bibitem{Lee:2009}
H.~J. Lee et~al.
\newblock X-ray thomson-scattering measurements of density and temperature in
  shock-compressed beryllium.
\newblock {\em Phys. Rev. Lett.}, 102:115001, Mar 2009.

\bibitem{PhysRevLett.109.265003}
S.~P. Regan et~al.
\newblock Inelastic x-ray scattering from shocked liquid deuterium.
\newblock {\em Phys. Rev. Lett.}, 109:265003, Dec 2012.

\bibitem{Fletcher:2015aa}
L.~B. Fletcher et~al.
\newblock Ultrabright x-ray laser scattering for dynamic warm dense matter
  physics.
\newblock {\em Nature Photonics}, 9:274--279, 03 2015.

\bibitem{PhysRevE.94.011202}
D.~Kraus et~al.
\newblock X-ray scattering measurements on imploding ch spheres at the national
  ignition facility.
\newblock {\em Phys. Rev. E}, 94:011202, Jul 2016.

\bibitem{Gregori:2007a}
G.~Gregori, A.~Ravasio, A.~H{\"o}ll, S.H. Glenzer, and S.J. Rose.
\newblock Derivation of the static structure factor in strongly coupled
  non-equilibrium plasmas for x-ray scattering studies.
\newblock {\em High Energy Density Physics}, 3(1):99 -- 108, 2007.

\bibitem{PhysRevE.74.026402}
G.~Gregori, S.~H. Glenzer, and O.~L. Landen.
\newblock Generalized x-ray scattering cross section from nonequilibrium
  plasmas.
\newblock {\em Phys. Rev. E}, 74:026402, Aug 2006.

\bibitem{Chihara:1987}
J.~Chihara.
\newblock Difference in x-ray scattering between metallic and non-metallic
  liquids due to conduction electrons.
\newblock {\em Journal of Physics F: Metal Physics}, 17(2):295, 1987.

\bibitem{Labrosse2010}
N.~Labrosse, P.~Heinzel, J.-C. Vial, T.~Kucera, S.~Parenti, S.~Gun{\'a}r,
  B.~Schmieder, and G.~Kilper.
\newblock Physics of solar prominences: I---spectral diagnostics and non-lte
  modelling.
\newblock {\em Space Science Reviews}, 151(4):243--332, Apr 2010.

\bibitem{Cottam:2002aa}
J.~Cottam, F.~Paerels, and M.~Mendez.
\newblock Gravitationally redshifted absorption lines in the x-ray burst
  spectra of a neutron star.
\newblock {\em Nature}, 420:51--54, 11 2002.

\bibitem{Bailey:2014aa}
J.~E. Bailey et~al.
\newblock A higher-than-predicted measurement of iron opacity at solar interior
  temperatures.
\newblock {\em Nature}, 517:56--59, 12 2014.

\bibitem{Hahn:2010}
David~W. Hahn and Nicol\'{o} Omenetto.
\newblock Laser-induced breakdown spectroscopy (libs), part i: Review of basic
  diagnostics and plasma--particle interactions: Still-challenging issues
  within the analytical plasma community.
\newblock {\em Appl. Spectrosc.}, 64(12):335A--366A, Dec 2010.

\bibitem{Liu:2013aa}
Xiaosong Liu, Dongdong Wang, Gao Liu, Venkat Srinivasan, Zhi Liu, Zahid
  Hussain, and Wanli Yang.
\newblock Distinct charge dynamics in battery electrodes revealed by in situ
  and operando soft x-ray spectroscopy.
\newblock {\em Nature Communications}, 4:2568, 10 2013.

\bibitem{PhysRevLett.100.167402}
Aron Walsh et~al.
\newblock Nature of the band gap of ${\mathrm{in}}_{2}{\mathrm{o}}_{3}$
  revealed by first-principles calculations and x-ray spectroscopy.
\newblock {\em Phys. Rev. Lett.}, 100:167402, Apr 2008.

\bibitem{Vinko2012}
S.~M. Vinko et~al.
\newblock {Creation and diagnosis of a solid-density plasma with an X-ray
  free-electron laser.}
\newblock {\em Nature}, 482(7383):59--62, 2012.

\bibitem{quincy:2018}
Q.~Y. van~den Berg et~al.
\newblock Clocking femtosecond collisional dynamics via resonant x-ray
  spectroscopy.
\newblock {\em Phys. Rev. Lett.}, 120:055002, Feb 2018.

\bibitem{PhysRevA.46.R1747}
R.~S. Marjoribanks, M.~C. Richardson, P.~A. Jaanimagi, and R.~Epstein.
\newblock Electron-temperature measurement in laser-produced plasmas by the
  ratio of isoelectronic line intensities.
\newblock {\em Phys. Rev. A}, 46:R1747--R1750, Aug 1992.

\bibitem{PhysRevLett.110.125001}
J.~Colgan et~al.
\newblock Exotic dense-matter states pumped by a relativistic laser plasma in
  the radiation-dominated regime.
\newblock {\em Phys. Rev. Lett.}, 110:125001, Mar 2013.

\bibitem{Hoarty2013}
D.~J. Hoarty et~al.
\newblock {Observations of the Effect of Ionization-Potential Depression in Hot
  Dense Plasma}.
\newblock {\em Phys. Rev. Lett.}, 110(26):265003, June 2013.

\bibitem{Regan:2002}
S.~P. Regan et~al.
\newblock Characterization of direct-drive-implosion core conditions on omega
  with time-resolved ar k-shell spectroscopy.
\newblock {\em Physics of Plasmas}, 9(4):1357--1365, 2002.

\bibitem{PhysRevLett.70.1263}
B.~A. Hammel, C.~J. Keane, M.~D. Cable, D.~R. Kania, J.~D. Kilkenny, R.~W. Lee,
  and R.~Pasha.
\newblock X-ray spectroscopic measurements of high densities and temperatures
  from indirectly driven inertial confinement fusion capsules.
\newblock {\em Phys. Rev. Lett.}, 70:1263--1266, Mar 1993.

\bibitem{Regan:2013}
S.~P. Regan et~al.
\newblock Hot-spot mix in ignition-scale inertial confinement fusion targets.
\newblock {\em Phys. Rev. Lett.}, 111:045001, 2013.

\bibitem{PhysRevLett.111.085004}
T.~Ma et~al.
\newblock Onset of hydrodynamic mix in high-velocity, highly compressed
  inertial confinement fusion implosions.
\newblock {\em Phys. Rev. Lett.}, 111:085004, Aug 2013.

\bibitem{Ciricosta:2017}
O.~Ciricosta et~al.
\newblock Simultaneous diagnosis of radial profiles and mix in nif
  ignition-scale implosions via x-ray spectroscopy.
\newblock {\em Physics of Plasmas}, 24(11):112703, 2017.

\bibitem{Lindl:2004}
J.~D. Lindl et~al.
\newblock The physics basis for ignition using indirect-drive targets on the
  national ignition facility.
\newblock {\em Phys. Plasmas}, 11(2):339--491, 2004.

\bibitem{Hammel:2010}
B.~Hammel et~al.
\newblock High-mode rayleigh-taylor growth in nif ignition capsules.
\newblock {\em High Energy Density Physics}, 6(2):171--178, 2010.

\bibitem{Scott:2001}
Howard~A. Scott.
\newblock Cretin---a radiative transfer capability for laboratory plasmas.
\newblock {\em Journal of Quantitative Spectroscopy and Radiative Transfer},
  71(2--6):689 -- 701, 2001.

\bibitem{Haan:2011}
S.~W. Haan et~al.
\newblock Point design targets, specifications, and requirements for the 2010
  ignition campaign on the national ignition facility.
\newblock {\em Phys. Plasmas}, 18(5):051001, 2011.

\bibitem{Ciricosta2012}
O.~Ciricosta et~al.
\newblock {Direct Measurements of the Ionization Potential Depression in a
  Dense Plasma}.
\newblock {\em Phys. Rev. Lett.}, 109:065002, 2012.

\bibitem{Ciricosta:2016aa}
O.~Ciricosta et~al.
\newblock Measurements of continuum lowering in solid-density plasmas created
  from elements and compounds.
\newblock {\em Nat. Commun.}, 7:11713, 05 2016.

\bibitem{Kozlowski:2016aa}
P.~M. Kozlowski, B.~J.~B. Crowley, D.~O. Gericke, S.~P. Regan, and G.~Gregori.
\newblock Theory of thomson scattering in inhomogeneous media.
\newblock {\em Scientific Reports}, 6:24283, 04 2016.

\bibitem{Crowley:2014aa}
B.J.B. Crowley and G.~Gregori.
\newblock Quantum theory of thomson scattering.
\newblock {\em High Energy Density Physics}, 13:55 -- 83, 2014.

\end{thebibliography}
\end{document}